\documentclass[12pt]{iopart}

\begin{document}

\title[Monomiality principle, Sheffer-type polynomials and ...]
{Monomiality principle, Sheffer-type polynomials and the normal ordering problem}
\author{K A Penson$^{a}$, P Blasiak$^{a,b}$, G Dattoli$^{c}$, G H E Duchamp$^{d}$, A Horzela$^{b}$
and A I Solomon$^{a,e}$\vspace{2mm}}

\address
{$^a$ Laboratoire de Physique Th\'eorique de la Mati\`{e}re Condens\'{e}e\\
Universit\'e Pierre et Marie Curie, CNRS UMR 7600\\
Tour 24 - 2i\`{e}me \'et., 4 pl. Jussieu, F 75252 Paris Cedex 05, France\vspace{2mm}}

\address
{$^b$ H. Niewodnicza\'nski Institute of
Nuclear Physics, Polish Academy of Sciences\\
ul. Eliasza-Radzikowskiego 152,  PL 31342 Krak\'ow, Poland\vspace{2mm}}

\address
{$^c$ ENEA, Dipartimento Innovazione, Divisione Fisica Applicata\\ 
Centro Ricerche Frascati, Via E. Fermi 45, I 00044 Frascati, Rome, Italy\vspace{2mm}}

\address{$^d$ Institut Galil\'ee, LIPN, CNRS UMR 7030\linebreak
99 Av. J.-B. Clement, F-93430 Villetaneuse, France\vspace{2mm}}

\address
{$^e$ The Open University, Physics and Astronomy Department\\
Milton Keynes MK7 6AA, United Kingdom\vspace{2mm}}

\eads{\linebreak \mailto{penson@lptl.jussieu.fr},
\mailto{ pawel.blasiak@ifj.edu.pl},
\mailto{ dattoli@frascati.enea.it},
\mailto{ gded@lipn-univ.paris13.fr},
\mailto{ andrzej.horzela@ifj.edu.pl},
\mailto{ a.i.solomon@open.ac.uk}}


\begin{abstract}
\\
We solve the boson normal ordering problem for
$\left(q(a^\dag)a+v(a^\dag)\right)^n$ with arbitrary functions
$q(x)$ and $v(x)$ and
integer $n$, where $a$ and $a^\dag$ are boson annihilation and
creation operators, satisfying $[a,a^\dag]=1$. This consequently provides the solution for the
exponential $e^{\lambda(q(a^\dag)a+v(a^\dag))}$ generalizing the
shift operator. In the course of these considerations we define
and explore the monomiality principle and find its
representations. We exploit the properties of Sheffer-type
polynomials which constitute the inherent structure of this
problem. In the end we give some examples illustrating the utility
of the method and point out the relation to combinatorial structures.
\end{abstract}


\section{Introduction}

In this work we are concerned with one mode boson creation
$a^\dag$ and annihilation $a$ operators satisfying the commutation relation
\begin{eqnarray}\label{HW}
[a,a^\dag]=1.
\end{eqnarray}
We consider the normal ordering problem of a specific class of operator
expressions.
The normally ordered form of a general function $F(a,a^{\dag})$,
denoted as ${\cal N}\left[F(a,a^{\dag})\right]\equiv F(a,a^{\dag})$, is defined by
moving all the annihilation operators $a$ to the
right, using the commutation relation Eq.(\ref{HW}). Additionally we define
the operation  $:\!G(a,a^{\dag})\!:$ which means normally order
$G(a,a^{\dag})$ {\it without} taking into account the commutation
relations. Using the latter operation
the normal ordering problem for $F(a,a^{\dag})$ is solved if we
find an operator $G(a,a^{\dag})$ for which 
$F(a,a^{\dag}) = \ :\!G(a,a^{\dag})\!:$ is satisfied.
\\
In 1974 J. Katriel \cite{Katriel} considered the normal ordering problem for powers of the number operator $N=a^\dag a$ and pointed out its connection to combinatorics. It can be written as
\begin{eqnarray}\label{Snk}
\left(a^\dag a\right)^n=\sum_{k=1}^n S(n,k) (a^\dag)^k a^k,
\end{eqnarray}
where the integers $S(n,k)$ are the so called \emph{Stirling
numbers} of the second kind counting the number of ways of putting 
$n$ different objects into $k$ identical containers (none left empty).
\\
In the coherent state representation $|z\rangle=e^{-|z|^2/2}\sum_{n=0}^\infty \frac{z^n}{\sqrt{n!}}|n\rangle$, where $a^\dag a |n\rangle =n|n\rangle $, $\langle n|n'\rangle =\delta_{n,n'}$ and $a |z\rangle =z|z\rangle $ \cite{Klauder}, we may write
\begin{eqnarray}
\langle z|\left(a^\dag a\right)^n|z\rangle=B(n,|z|^2),
\end{eqnarray}
where $B(n,x)$ are so called (exponential) \emph{Bell polynomials} \cite{Comtet}
\begin{eqnarray}\label{B}
B(n,x)=\sum_{k=1}^n S(n,k) x^k.
\end{eqnarray}
Further development of this idea (see {\it e.g.} \cite{Katriel2002},\cite{BlasiakPhD}) provides the normally ordered expression for the exponential 
\begin{eqnarray}\label{eaa}
e^{\lambda a^\dag a}={\cal N}\left[e^{\lambda a^\dag
a}\right]\equiv\ :e^{a^\dag a(e^\lambda-1)}\ :.
\end{eqnarray}
Here we shall extend these results in a
very particular direction. We consider operators linear in
annihilation $a$ or creation $a^\dag$ operators. More
specifically, we consider  operators which, say for linearity in
$a$, have the form
$q(a^\dag)a+v(a^\dag)$,
where $q(x)$ and $v(x)$ are arbitrary functions.
We shall find the normally ordered form of the $n$-th power
$(q(a^\dag)a+v(a^\dag))^n$
and then of the exponential
$e^{\lambda (q(a^\dag)a+v(a^\dag))}$.
This is a far reaching generalization of the results of
\cite{Mikhailov1983},\cite{Mikhailov1985},\cite{Katriel1983} where a
special case of the operator $a^\dag a+a^r$ was considered.
\\
In this approach we use methods which are based on the
monomiality principle \cite{BlasiakJPA2005}. First, using the methods of
umbral calculus, we find a wide
class of representations of the canonical commutation relation
Eq.(\ref{HW}) in the space of polynomials. 
This establishes the link with Sheffer-type
polynomials. Next the specific matrix elements of the above
operators are derived and thereafter, with the help of coherent state
representation, extended to the general form. Finally we obtain
normally ordered expressions for these operators. It turns out that
the exponential generating functions in the case of linear
dependence on the annihilation (or creation) operator are of
Sheffer-type, and  that assures their convergence.
In the end we give some examples with special emphasis put on their
Sheffer-type origin and point out the relation to combinatorial structures.

\section{Monomiality principle}

Here we introduce the concept of monomiality which arises from the
action of the multiplication and derivative operators on
monomials. Next we provide a wide class of representations of that
property in the framework of Sheffer-type polynomials. Finally we
establish the correspondence to the occupation number
representation.

\subsection{Definition and general properties}

Let us consider the Heisenberg-Weyl algebra satisfying the
commutation relation
\begin{eqnarray}\label{PM}
[P,M]=1.
\end{eqnarray}
The simplest
representation of Eq.(\ref{PM}) is by the derivative $P=D=\frac{d}{dx}$ and
multiplication $M=X$ operators acting in the space of
polynomials ($[D,X]=1$). They are defined by their action on monomials 
\begin{eqnarray}\label{DXMono}
\begin{array}{lcl}
X x^n&=&x^{n+1},\\
D x^n&=&nx^{n-1}.
\end{array}
\end{eqnarray}
and subsequently on polynomials and formal power series.
\\
Here we extend this framework. Suppose one wants to find the representations of Eq.(\ref{PM})
such that the action of $M$ and $P$ on certain polynomials
$s_n(x)$ is analogous to the action of $X$ and $D$ on monomials.
More specifically one searches for $M$ and $P$ and
associated polynomials $s_n(x)$ (of degree $n$, $n=0,1,2,...$)
which satisfy
\begin{eqnarray}\label{Monomiality}
\begin{array}{l}
Ms_n(x)=s_{n+1}(x),\\
Ps_n(x)=n\ s_{n-1}(x).
\end{array}
\end{eqnarray}
The rule embodied in Eq.(\ref{Monomiality}) is called the {\em
monomiality principle}. The polynomials $s_n(x)$ are then called
{\it quasi-monomials} with respect to operators $M$ and $P$. These
operators can be immediately recognized as raising and lowering
operators acting on the $s_n(x)$'s.
\\
The definition Eq.(\ref{Monomiality}) implies some general
properties besides fulfilling commutation relation of Eq.(\ref{PM}). First the operators $M$ and $P$ obviously satisfy
Eq.(\ref{PM}). Further consequence of Eq.(\ref{Monomiality}) is
the eigenproperty of $MP$
\begin{eqnarray}
MPs_n(x)=ns_n(x).
\end{eqnarray}
The polynomials $s_n(x)$ may be obtained through the action of
$M^n$ on $s_0$
\begin{eqnarray}\label{Mn}
s_n(x)=M^ns_0
\end{eqnarray}
and consequently the exponential generating function of $s_n(x)$'s
is
\begin{eqnarray}\label{G}
G(\lambda,x)\equiv \sum_{n=0}^\infty
s_n(x)\frac{\lambda^n}{n!}=e^{\lambda M}s_0.
\end{eqnarray}
Also, if we write the quasimonomial $s_n(x)$ explicitly as
\begin{eqnarray}\label{S00}
s_n(x)=\sum_{k=0}^n s_{n,k}\ x^k,
\end{eqnarray}
then
\begin{eqnarray}\label{SX}
s_n(x)=\left(\sum_{k=0}^n s_{n,k}\ X^k\right)1.
\end{eqnarray}
Several types of such polynomial sequences were studied recently
using this monomiality principle
\cite{DattoliNuovoCim},\cite{Dattoli1997},\cite{Dattoli1999},\cite{Dattoli2001},\cite{Cesarano2000}.

\subsection{Monomiality principle representations: Sheffer-type polynomials}
\label{representations}
Here we  show that if $s_n(x)$ are of \emph{Sheffer-type} then it
is possible to give explicit representations of $M$ and $P$.
Conversely, if  $M=M(X,D)$ and $P=P(D)$ then $s_n(x)$ of
Eq.(\ref{Monomiality}) are necessarily of Sheffer-type.
\\
Properties of Sheffer-type polynomials are naturally handled
within the so called {\it umbral calculus}
\cite{Roman},\cite{Rota},\cite{DiBucchianico}. 
They are usually defined through their exponential generating function 
\begin{eqnarray}\label{AB}
G(\lambda,x)=\sum_{n=0}^\infty
s_n(x)\frac{\lambda^n}{n!}=A(\lambda)e^{xB(\lambda)}
\end{eqnarray}
with some (formal) power series $A(\lambda)=\sum_{n=0}^\infty a_n \frac{\lambda^n}{n!}$ and $B(\lambda)=\sum_{n=0}^\infty b_n \frac{\lambda^n}{n!}$
such that $b_0=0$, $b_1\neq0$ and $a_0\neq0$.
\\
Here we focus on their ladder structure aspect and recall the following equivalent definition. Suppose we have a polynomial sequence $s_n(x)$,
$n=0,1,2,...$ ($s_n(x)$ being a polynomial of degree $n$). It is
called of a Sheffer A-type zero \cite{Sheffer},\cite{Rainville}
(which we shall call Sheffer-type) if there exists a function
$f(x)$ such that
\begin{eqnarray}\label{S0}
f(D)s_n(x)=ns_{n-1}(x).
\end{eqnarray}
Operator $f(D)$ plays the role of the lowering operator. This
characterization is not unique, {\it i.e.} there are many
Sheffer-type sequences $s_n(x)$ satisfying Eq.(\ref{S0}) for given
$f(x)$. We can further classify them by postulating the existence
of the associated raising operator. A general theorem
\cite{Roman},\cite{Cheikh} states that a polynomial sequence
$s_n(x)$ satisfying the monomiality principle
Eq.(\ref{Monomiality}) with an operator $P$ given as a function of
the derivative operator only $P=P(D)$ is {\it uniquely}
determined by two (formal) power series $f(x)=\sum_{n=0}^\infty f_n \frac{\lambda^n}{n!}$ and $g(x)=\sum_{n=0}^\infty g_n \frac{\lambda^n}{n!}$ such that $f_0=0$,
$f_1\neq0$ and $g_0\neq 0$. The exponential generating
function of $s_n(x)$ is then equal to
\begin{eqnarray}\label{egf}
G(\lambda,x)=\sum_{n=0}^\infty
s_n(x)\frac{\lambda^n}{n!}=\frac{1}{g(f^{-1}(\lambda))}\
e^{xf^{-1}(\lambda)},
\end{eqnarray}
and their associated raising and lowering operators of
Eq.(\ref{Monomiality}) are given by
\begin{eqnarray}\label{pm}
\begin{array}{l}
P=f(D),\\
M=\left[X-\frac{g'(D)}{g(D)}\right]\frac{1}{f'(D)}\ .
\end{array}
\end{eqnarray}
Observe that $X$ enters $M$ only linearly and the order of $X$ and $D$ in $M(X,D)$ matters.
\\
By direct calculation one may check that any pair $M$, $P$ from
Eq.(\ref{pm}) automatically satisfies Eq.(\ref{PM}). The detailed
proof can be found in \cite{Roman},\cite{Cheikh}.
\\
Here are some examples we have  obtained of representations of the
monomiality principle Eq.(\ref{Monomiality}) and their associated
Sheffer-type polynomials:

\begin{itemize}
\item[a)]{$M(X,D)=2X-D$, \ \ \ \ $P(D)=\frac{1}{2}D$,

$s_n(x)=H_n(x)$ - Hermite polynomials;

$G(\lambda,x)=e^{2\lambda x-\lambda^2}$.}

\item[b)]{$M(X,D)=-XD^2+(2X-1)D-X-1$, \ \ \ \ $P(D)=-\sum_{n=1}^\infty D^n$,

$s_n(x)=n!L_n(x)$ - where $L_n(x)$ are  Laguerre polynomials;

$G(\lambda,x)=\frac{1}{1-\lambda}e^{x\frac{\lambda}{\lambda -
1}}$.}

\item[c)]{$M(X,D)=X\frac{1}{1-D}$, \ \ \ \ $P(D)=-\frac{1}{2}D^2+D$,

$s_n(x)=P_n(x)$ - Bessel polynomials \cite{Grosswald};

$G(\lambda,x)=e^{x(1-\sqrt{1-2\lambda})}$.}

\item[d)]{$M(X,D)=X(1+D)$, \ \ \ \ $P(D)=\ln(1+D)$,

$s_n(x)=B_n(x)$ - (exponential) Bell polynomials;

$G(\lambda,x)=e^{x(e^\lambda-1)}$.}

\item[e)]{$M(X,D)=Xe^{-D}$, \ \ \ \ $P(D)=e^D-1$,

$s_n(x)=x^{\underline{n}}$ - the lower factorial
polynomials \cite{Turbiner};

$G(\lambda,x)=e^{x\ln (1+\lambda )}$.\vspace{2mm}}

\item[f)]{$M(X,D)=(X-\tan (D))\cos^2(D)$, \ \ \ \ $P(D)=\tan (D)$,

$s_n(x)=R_n(x)$ - Hahn polynomials \cite{BenderHahn};

$G(\lambda,x)=\frac{1}{\sqrt{1+\lambda^2}}e^{x\arctan(\lambda)}$.}

\item[g)]{$M(X,D)=X\frac{1+W_L(D)}{W_L(D)}D$, \ \ \ \ $P(D)=W_L (D)$,

where $W_L(x)$ is the Lambert $W$ function \cite{KnuthW};

$s_n(x)=I_n(x)$ - the idempotent polynomials \cite{Comtet};

$G(\lambda,x)=e^{x\lambda e^\lambda}$.}
\end{itemize}

\subsection{Monomiality vs Fock space representations}
\label{correspondence}
We have already called operators $M$ and $P$ satisfying
Eq.(\ref{Monomiality}) the rising and lowering operators. Indeed,
their action rises and lowers the index $n$ of the quasimonomial
$s_n(x)$. This resembles the property of creation $a^\dag$ and
annihilation $a$ operators in the Fock space given by
\begin{eqnarray}
\begin{array}{rcr}
a |n\rangle&=&\sqrt{n}\ |n-1\rangle,\\
a^\dag|n\rangle&=&\sqrt{n+1}\ |n+1\rangle.
\end{array}
\end{eqnarray}
These relations are almost the same as Eq.(\ref{Monomiality}).
There is a difference in coefficients. To make them analogous it
is convenient to redefine the number states $|n\rangle$ as
\begin{eqnarray}
\widetilde{|n\rangle}=\sqrt{n!}\ |n\rangle.
\end{eqnarray}
(Note that $\widetilde{|0\rangle}\equiv|0\rangle$).
\\
Then the creation and annihilation operators act as
\begin{eqnarray}
\begin{array}{rcl}
a \widetilde{|n\rangle}&=&n\ \widetilde{|n-1\rangle},\\
a^\dag\widetilde{|n\rangle}&=&\widetilde{|n+1\rangle}.
\end{array}
\end{eqnarray}
Now this exactly mirrors the relation of
Eq.(\ref{Monomiality}). So, we make the correspondence
\begin{eqnarray}
\begin{array}{ccl}
P&\ \ \longleftrightarrow \ \ &a \\
M&\ \ \longleftrightarrow \ \ &a^\dag\\
s_n(x)&\ \ \longleftrightarrow \ \ &\widetilde{|n\rangle}\ ,\ \ \ \ \ n=0,1,2,...\ \ . \\
\end{array}
\end{eqnarray}
We note that this identification is purely algebraic, {\it i.e.}
we are concerned here only with the commutation relation
Eqs.(\ref{HW}) or (\ref{PM}). We neither impose the
scalar product in the space of polynomials nor consider the
conjugacy properties of the operators. These properties are
irrelevant for our proceeding discussion. We note only that they
may be rigorously introduced, see {\it e.g.} \cite{Roman}.

\section{Normal ordering via monomiality}

In this section we shall exploit the correspondence of Section
\ref{correspondence} to obtain the normally ordered expression of
powers and exponential of the operators $a^\dag q(a)+v(a)$ and (by
the conjugacy property) $q(a^\dag)a+v(a^\dag)$. 
To this end we shall apply the
results of Section \ref{representations} to calculate some
specific coherent state matrix elements of operators in question
and then, through the exponential mapping property, we shall extend
them to general matrix elements. In conclusion we shall also comment on
other forms of linear dependence on $a$ or $a^\dag$.
\\
We use the correspondence of Section \ref{correspondence}
cast in the simplest form for $M=X$, $P=D$ and $s_n(x)=x^n$, {\it
i.e.}
\begin{eqnarray}\label{XDa}
\begin{array}{ccl}
D&\ \ \longleftrightarrow \ \ &a, \\
X&\ \ \longleftrightarrow \ \ &a^\dag ,\\
x^n&\ \ \longleftrightarrow \ \ &\widetilde{|n\rangle}\ ,\ \ \ \ \ n=0,1,2,...\ \ . \\
\end{array}
\end{eqnarray}
Now, recall the representation Eq.(\ref{pm}) of operators $M$
and $P$ in terms of $X$ and $D$. Applying the correspondence of
Eq.(\ref{XDa}) it takes the form
\begin{eqnarray}\label{PMa}
\begin{array}{l}
P(a)=f(a),\\
M(a,a^\dag)=\left[a^\dag-\frac{g'(a)}{g(a)}\right]\frac{1}{f'(a)}\ .
\end{array}
\end{eqnarray}
From Eqs.(\ref{Mn}),(\ref{S00}) and (\ref{SX}) we get
\begin{eqnarray}
\left[M(a,a^\dag)\right]^n|0\rangle=\sum_{k=0}^n s_{n,k}\ (a^\dag)^k|0\rangle.
\end{eqnarray}
In the coherent state representation it yields
\begin{eqnarray}\label{M0}
\langle z|\left[M(a,a^\dag)\right]^n|0\rangle=s_n(z^*)\langle z|0\rangle.
\end{eqnarray}
Exponentiating $M(a,a^\dag)$ and using Eq.(\ref{egf}) we obtain
\begin{eqnarray}\label{EM0}
\langle z|e^{\lambda M(a,a^\dag)}|0\rangle=\frac{1}{g(f^{-1}(\lambda))}\ e^{z^*f^{-1}(\lambda)}\langle z|0\rangle.
\end{eqnarray}
By the same token one obtains closed form expressions for the
following matrix elements ( $|l\rangle$ is the $l$-th number
state, $l=0,1,2,...$)
\begin{eqnarray}\label{Ml}
\langle z|\left[M(a,a^\dag)\right]^n|l\rangle=\frac{1}{\sqrt{l!}}s_{n+l}(z^*)\langle z|0\rangle,
\end{eqnarray}
and further
\begin{eqnarray}\label{El}
\langle z|e^{\lambda M(a,a^\dag)}|l\rangle=\frac{1}{\sqrt{l!}}
\frac{d^l}{d\lambda^l}\left[\frac{1}{g(f^{-1}(\lambda))}\ e^{z^*f^{-1}(\lambda)}\right]\langle z|0\rangle.
\end{eqnarray}
Observe that in both Eqs.(\ref{M0}) and (\ref{Ml}) we obtain
Sheffer-type polynomials (modulus coherent states overlapping
factor $\langle z|0\rangle$). Also Eqs.(\ref{EM0}) and (\ref{El})
reveal that property through the Sheffer-type generating
function. This connection will be explored in detail in Section
\ref{sheffernormal}.
\\
The result of Eq.(\ref{EM0}) can be further extended to the
general matrix element $\langle z|e^{\lambda
M(a,a^\dag)}|z'\rangle$. To this end recall the property $|z\rangle=e^{-\frac{|z|^2}{2}}e^{za^\dag}|0\rangle$ and
write
\begin{eqnarray}\nonumber
\langle z|e^{\lambda M(a,a^\dag)}|z'\rangle&=&e^{-\frac{1}{2}|z'|^2}\langle z|e^{\lambda M(a,a^\dag)}e^{z'a^\dag}|0\rangle\\\nonumber
&=&e^{-\frac{1}{2}|z'|^2}\langle
z|e^{z'a^\dag}e^{-z'a^\dag}e^{\lambda
M(a,a^\dag)}e^{z'a^\dag}|0\rangle
\\\nonumber &=&e^{z^*z'-\frac{1}{2}|z'|^2}\langle z|e^{-z'a^\dag}e^{\lambda M(a,a^\dag)}e^{z'a^\dag}|0\rangle.
\end{eqnarray}
Next, using the exponential mapping formula $e^{-xa^\dag}F(a,a^\dag)e^{xa^\dag}=F(a+x,a^\dag)$ (see \cite{Klauder},\cite{Louisell}) we
arrive at
\begin{eqnarray}\nonumber
\langle z|e^{\lambda M(a,a^\dag)}|z'\rangle&=&e^{z^*z'-\frac{1}{2}|z'|^2}\langle z|e^{\lambda M(a+z',a^\dag)}|0\rangle\\\nonumber
&=&e^{z^*z'-\frac{1}{2}|z'|^2}\langle
z|e^{\lambda\left(a^\dag-\frac{g'(a+z')}{g(a+z')}\right)
\frac{1}{f'(a+z')}}|0\rangle.
\end{eqnarray}
Now we are almost ready to apply Eq.(\ref{EM0}) to evaluate the
matrix element on the r.h.s. of the above equation. Before doing
so we have to appropriately redefine functions $f(x)$ and $g(x)$
in the following way ($z'$ - fixed)
\begin{eqnarray}\nonumber
    f(x)&\to& \tilde{f}(x)=f(x+z')-f(z'),\\\nonumber
    g(x)&\to& \tilde{g}(x)=g(x+z')/g(z').
\end{eqnarray}
Then $\tilde{f}(0)=0$, $\tilde{f}'(0)\neq 0$ and $\tilde{g}(0)=1$
as required by the Sheffer property for $\tilde{f}(x)$ and
$\tilde{g}(x)$. Observe that these conditions are not fulfilled by
$f(x+z')$ and $g(x+z')$. This step imposes (analytical)
constraints on $z'$, {\it i.e.} it is valid whenever
$\tilde{f}'(z')\neq 0$ (although, we note that for formal power
series approach this does not present any difficulty). Now we can
write
\begin{eqnarray}\nonumber
\langle z|e^{\lambda\left(a^\dag-\frac{g'(a+z')}{g(a+z')}\right)
\frac{1}{f'(a+z')}}|0\rangle&=&\langle z|e^{\lambda\left(a^\dag-\frac{\tilde{g}'(a)}{\tilde{g}(a)}\right)
\frac{1}{\tilde{f}'(a)}}|0\rangle\\\nonumber
&=&\frac{1}{\tilde{g}(\tilde{f}^{-1}(\lambda))}\
e^{z^*\tilde{f}^{-1}(\lambda)}\langle z|0\rangle.
\end{eqnarray}
By going back to the initial functions $f(x)$ and $g(x)$ this
readily gives the final result
\begin{eqnarray}\label{Ze}
\langle z|e^{\lambda M(a,a^\dag)}|z'\rangle
=\frac{g(z')}{g(f^{-1}(\lambda+f(z')))}e^{z^*[f^{-1}(\lambda+f(z'))-z']}\langle
z|z'\rangle,
\end{eqnarray}
where $\langle
z|z'\rangle=e^{z^*z'-\frac{1}{2}|z^{'}|^2-\frac{1}{2}|z|^2}$ is
the coherent states overlapping factor.
\\
To obtain the normally ordered form of $e^{\lambda M(a,a^\dag)}$ we
apply the crucial property of the coherent state representation 
$\langle z|{F}(a,a^{\dag})|z'\rangle = \langle z|z'\rangle\ G(z^*,z')
\Longrightarrow F(a,a^{\dag}) =\ :G(a^{\dag},a):$
(see \cite{Klauder}). Then Eq.(\ref{Ze}) provides the
central result
\begin{eqnarray}\label{Normal}
e^{\lambda M(a,a^\dag)}=\
:e^{a^\dag[f^{-1}(\lambda+f(a))-a]}\frac{g(a)}{g(f^{-1}(\lambda+f(a)))}:\
.
\end{eqnarray}
Let us point out again that $a^\dag$ appears linearly in
$M(a,a^\dag)$, see Eq.(\ref{PMa}).
\\
For simplicity we put
\begin{eqnarray}\nonumber
    q(x)&=&\frac{1}{f'(x)},\\\nonumber
    v(x)&=&\frac{g'(x)}{g(x)}\frac{1}{f'(x)},
\end{eqnarray}
and define
\begin{eqnarray}\nonumber
    T(\lambda,x)&=&f^{-1}(\lambda+f(x)),\\\nonumber
    G(\lambda,x)&=&\frac{g(x)}{g(T(\lambda,x))}.
\end{eqnarray}
This allows us to rewrite the main normal ordering formula of
Eq.(\ref{Normal}) as
\begin{eqnarray}\label{normal}
e^{\lambda \left(a^\dag q(a)+v(a)\right)}=
\ :e^{a^\dag[T(\lambda,a)-a]}\ G(\lambda,a):
\end{eqnarray}
where the functions $T(\lambda,x)$ and $G(\lambda,x)$ fulfill the
following differential equations
\begin{eqnarray}\label{t}
\frac{\partial T(\lambda,x)}{\partial\lambda} =  q(T(\lambda,x))\ , &~~~~~~~~~~~~~~&T(0,x) = x\ ,
\end{eqnarray}
\begin{eqnarray}
\label{g}
\frac{\partial G(\lambda,x)}{\partial\lambda} =  v(T(\lambda,x))\cdot G(\lambda,x)\ , &~~~~&G(0,x) = 1\ .
\end{eqnarray}
Eq.(\ref{normal}) in the coherent state representation takes the form
\begin{eqnarray}\label{zQz}
\langle z'|e^{\lambda \left(a^\dag q(a)+v(a)\right)}|z\rangle=\langle z'|z\rangle\
e^{z'^*[T(\lambda,z)-z]} G(\lambda,z).
\end{eqnarray}
We conclude by making a comment on other possible forms of linear
dependence on $a$ or $a^\dag$.
\\
By hermitian conjugation of Eq.(\ref{normal}) we obtain the
expression for the normal form of $e^{\lambda
\left(q(a^\dag)a+v(a)\right)}$. This amounts to the formula
\begin{eqnarray}\label{aqa}
e^{\lambda \left(q(a^\dag)a+v(a)\right)}=
\ :G(\lambda,a^\dag)e^{[T(\lambda,a^\dag)-a^\dag]a}:\
\end{eqnarray}
with the same differential equations Eqs.(\ref{t}) and (\ref{g})
for functions $T(\lambda,x)$ and $G(\lambda,x)$. In the coherent
state representation it yields
\begin{eqnarray}\label{zqz}
\langle z'|e^{\lambda \left(q(a^\dag)a+v(a)\right)}|z\rangle=\langle z'|z\rangle\
 G(\lambda,z'^*)e^{[T(\lambda,z'^*)-z'^*]z}
\end{eqnarray}
We also note that all other operators linearly dependent on $a$ or
$a^\dag$ may be written in just considered forms using
$[a,F(a,a^\dag)]=\frac{\partial}{\partial a^\dag}\ F(a,a^\dag)$ which yields
$aq(a^\dag)+v(a^\dag)=q(a^\dag)a+q'(a^\dag)+v(a^\dag)$ and
$q(a)a^\dag+v(a)=a^\dag q(a)+q'(a)+v(a)$.
\\
Observe that from analytical point of view certain limitations 
on the domains of $z$, $z'$ and $\lambda$ should be put in some specific cases 
(locally around zero all the formulas hold true). Also we point out
the fact that functions $q(x)$ and $v(x)$ (or equivalently $f(x)$ and $g(x)$)
may be taken as the formal power series. 
\\
In the end we note
that the reverse process, {\it i.e.} derivation of the normally
ordered form from the substitution theorem, is also possible, see
\cite{BlasiakPLA2005}.

\section{Sheffer-type polynomials and normal ordering: Examples}
\label{ShefferExamples}

We now proceed to examples putting special emphasis on their
Sheffer-type origin.

\subsection{Examples}

We start with enumerating some examples of the evaluation of the
coherent state matrix elements of Eqs.(\ref{M0}) and (\ref{zQz}). We
choose the $M(a,a^\dag)$'s as in the list a) -  g) in Section
\ref{representations}:

\begin{itemize}

\item[a)]{$\langle z|(-a+2a^\dag)^n|0\rangle =H_n(z^*)\langle z|0\rangle$, Hermite polynomials;

$\langle z|e^{\lambda (-a+2a^\dag)}|z'\rangle=e^{\lambda(2 z^*-
z')-\lambda^2}\langle z|z'\rangle$.}

\item[b)]{$\langle z|\left[-a^\dag a^2+(2a^\dag-1)a-a^\dag+1\right]^n|0\rangle =n!L_{n}(z^*)\langle z|0\rangle$,

Laguerre polynomials;

$\langle z|e^{\lambda \left[-a^\dag
a^2+(2a^\dag-1)a-a^\dag+1\right]}|z'\rangle
=\frac{z'^2 -\lambda z' + 1}{(1 - z')(1-\lambda(z'-1))
}e^{z^*\lambda\frac{(1-z')^2}{(\lambda(1-z') - 1)}} \langle
z|z'\rangle$.}

\item[c)]{$\langle z|\left(a^\dag\frac{1}{1-a}\right)^n|0\rangle =P_n(z^*)\langle z|0\rangle$, Bessel polynomials;

$\langle z|e^{\lambda \left(a^\dag\frac{1}{1-a}\right)}|z'\rangle
=e^{z^*[1-\sqrt{1-2(\lambda+z'-\frac{1}{2}z'^2)}-z']}\langle
z|z'\rangle$.}

\item[d)]{$\langle z|(a^\dag a+a^\dag)^n|0\rangle =B_n(z^*)\langle z|0\rangle$, Bell polynomials;

$\langle z|e^{\lambda (a^\dag
a+a^\dag)}|z'\rangle=e^{z^*(z'+1)(e^\lambda-1)}\langle
z|z'\rangle$.}

\item[e)]{$\langle z|(a^\dag e^{-a})^n|0\rangle =(z^*)^{\underline{n}}\langle z|0\rangle$, the lower factorial polynomials;

$\langle z|e^{\lambda (a^\dag
e^{-a})}|z'\rangle=e^{z^*[\ln(e^{z'}+\lambda)-z']}\langle
z|z'\rangle$.}

\item[f)]{$\langle z|\left[(a^\dag-\tan (a))\cos^2(a)\right]^n|0\rangle =R_n(z^*)\langle z|0\rangle$, Hahn polynomials;

$\langle z|e^{\lambda (a^\dag-\tan (a))\cos^2(a)}|z'\rangle
=\frac{\cos[\arctan
(\lambda+\tan(z'))]}{cos(z')}e^{z^*[\arctan(\lambda\tan(z'))-z']}\langle
z|z'\rangle$.}

\item[g)]{$\langle z|\left[a^\dag\frac{1+W_L(a)}{W_L(a)}a\right]^n|0\rangle =I_n(z^*)\langle z|0\rangle$, the idempotent polynomials;

$\langle z|e^{\lambda
a^\dag\frac{1+W_L(a)}{W_L(a)}a}|z'\rangle=e^{z^*[\lambda
e^{\lambda+W_L(z')}+z'(e^\lambda-1)]}\langle z|z'\rangle$.}
\end{itemize}
Note that for $z'=0$ we obtain the exponential generating
functions of appropriate polynomials multiplied by the coherent
states overlapping factor $\langle z|0\rangle$, see Eq.(\ref{Ze}).
\\
These examples show how the Sheffer-type polynomials and their
exponential generating functions arise in the coherent state
representation. This generic structure is the consequence of
Eqs.(\ref{M0}) and (\ref{Ze}) or in general Eqs.(\ref{zQz}) or
(\ref{zqz}) and it will be investigated in more detail now.
Afterwords we shall provide more examples of combinatorial origin.

\subsection{Sheffer polynomials and normal ordering}
\label{sheffernormal}

First recall the definition of the family of Sheffer-type
polynomials $s_n(z)$ defined through
the exponential generating function (see Eq.(\ref{AB})) as
\begin{eqnarray}\label{egfsheffer}
G(\lambda,x)=\sum_{n=0}^\infty
s_n(z)\frac{\lambda^n}{n!}=A(\lambda)\ e^{zB(\lambda)}
\end{eqnarray}
where functions $A(\lambda)$ and $B(\lambda)$ satisfy: $A(0)\neq
0$ and $B(0)=0$, $B'(0)\neq 0$.
\\
Returning to normal ordering, recall that the coherent state
expectation value of Eq.(\ref{aqa}) is given by Eq.(\ref{zqz}).
When one
\underline{fixes} $z'$ and takes $\lambda$ and $z$ as indeterminates, then the r.h.s. of Eq.(\ref{zqz})
may be read off as an exponential generating function of
Sheffer-type polynomials defined by Eq.(\ref{egfsheffer}). The
correspondence is given by
\begin{eqnarray}
\label{AB1}
A(\lambda)=g(\lambda,z'^*),\\
\label{AB2}
B(\lambda)=\left[T(\lambda,z'^*)-z'^*\right].
\end{eqnarray}
This allows us to make the statement that the coherent state
expectation value $\langle z'|...|z\rangle$ of the operator
$\exp\left[\lambda(q(a^\dag)a+v(a^\dag))\right]$ for any fixed
$z'$ yields (up to the overlapping factor $\langle z'|z\rangle$)
the exponential generating function of a certain sequence of
Sheffer-type polynomials in the variable $z$ given by
Eqs.(\ref{AB1}) and (\ref{AB2}). The above construction
establishes the connection between the coherent state
representation of the operator
$\exp\left[\lambda(q(a^\dag)a+v(a^\dag))\right]$ and a family of
Sheffer-type polynomials $s^{(q,v)}_n(z)$ related to $q$ and $v$
through
\begin{eqnarray}
\label{shefferseq}
\langle z^{\prime}|e^{\lambda\left[q(a^{\dag})a + v(a^{\dag})\right]}|z\rangle =
\langle z^{\prime}|z\rangle\left( 1+\sum_{n=1}^\infty s_n^{(q,v)}(z)\frac{\lambda^n}{n!}\right),
\end{eqnarray}
where explicitly (again for $z'$ fixed):
\begin{equation}
\label{shefferseq2}
\begin{array}{rcl}
s_n^{(q,v)}(z)=\langle z^{\prime}|z\rangle^{-1}\langle
z^{\prime}|\left[q(a^{\dag})a + v(a^{\dag})\right]^n|z\rangle.
\end{array}
\end{equation}
We observe that Eq.(\ref{shefferseq2}) is an extension of the seminal formula of Katriel
\cite{Katriel},\cite{Katriel2000} where $v(x)=0$ and $q(x)=x$. The Sheffer-type polynomials are
in this case Bell polynomials expressible through the Stirling
numbers of the second kind Eq.(\ref{B}).
\\
Having established relations leading from the normal ordering
problem to Sheffer-type polynomials we may consider the reverse
approach. Indeed, it turns out that for any Sheffer-type sequence
generated by $A(\lambda)$ and $B(\lambda)$ one can find functions
$q(x)$ and $v(x)$ such that the coherent state expectation value
$\langle
z'|\exp\left[\lambda(q(a^\dag)a+v(a^\dag))\right]|z\rangle$
results in a corresponding exponential generating function of
Eq.(\ref{egfsheffer}) in indeterminates $z$ and $\lambda$ (up to
the overlapping factor $\langle z'|z\rangle$ and $z'$ fixed).
Appropriate formulas can be derived from Eqs.(\ref{AB1}) and
(\ref{AB2}) by substitution into Eqs.(\ref{t}) and (\ref{g}):
\begin{eqnarray}\label{111}
q(x)&=&B'(B^{-1}(x-z'^*)),\\\label{2} v(x)&=&\frac{A'(B^{-1
}(x-z'^*))}{A(B^{-1 }(x-z'^*))}.
\end{eqnarray}
\noindent One can check that this choice of $q(x)$ and $v(x)$, if inserted into
Eqs. (\ref{t}) and (\ref{g}), results in
\begin{eqnarray}\label{3}
T(\lambda,x)&=&B(\lambda+B^{-1 }(x-z'^*))+z'^*,\\\label{4}
g(\lambda,x)&=&\frac{A(\lambda+B^{-1}(x-z'^*))}{A(B^{-1}(x-z'^*))},
\end{eqnarray}
which reproduce
\begin{eqnarray}\label{zABz}
\langle z^{\prime}|e^{\lambda\left[q(a^{\dag})a + v(a^{\dag})\right]}|z\rangle = \langle z^{\prime}|z\rangle A(\lambda)e^{zB(\lambda)}.
\end{eqnarray}
\noindent The result summarized in Eqs.(\ref{AB1}) and (\ref{AB2}) and in their 'dual' forms Eqs.(\ref{111})-(\ref{4})
provide us with a considerable flexibility in conceiving and
analyzing a large number of examples.

\subsection{Combinatorial examples}
In this section we will work out examples illustrating how the
exponential generating function $G(\lambda)=\sum_{n=0}^\infty
a_n\frac{x^n}{n!}$ of certain combinatorial sequences
$(a_n)_{n=0}^\infty$  appear naturally in the context of boson
normal ordering. To this end we shall assume specific forms of
$q(x)$ and $v(x)$ thus specifying the operator that we
exponentiate. We then give solutions to Eqs.(\ref{t}) and
(\ref{g}) and subsequently through Eqs.(\ref{AB1}) and (\ref{AB2})
we write the exponential generating function of combinatorial
sequences whose interpretation will be given.

\begin{itemize}
\item[a)]{ Choose $q(x)=x^r$, $r>1$ (integer), $v(x)=0$ (which implies
$g(\lambda,x)=1$). Then $T(\lambda,x) = x\left[1 - \lambda(r -
1)x^{r-1}\right]^{\frac{1}{1-r}}$. This gives
\begin{eqnarray}\nonumber
{\cal N}\left[e^{\lambda (a^\dag)^ra}\right] \equiv
\ : \exp\left[\left(\frac{a^\dag}{\left(1 - \lambda(r - 1)(a^\dag)^{r-1}\right)^{\frac{1}{r-1}}}-1\right)a\right]:\
\end{eqnarray}
as the normally ordered form. Now we take $z^{'}=1$ in
Eqs.(\ref{AB1}) and (\ref{AB2}) and from Eq.(\ref{zABz}) we obtain
\begin{eqnarray}\nonumber
\langle 1|z\rangle^{-1}\langle 1|e^{\lambda (a^\dag)^ra}|z\rangle  =\ \exp\left[z\left(\frac{1}{\left(1 - \lambda(r - 1)\right)^{\frac{1}{r-1}}}-1\right)\right]\ ,
\end{eqnarray}
which for $z=1$ generates the following sequences:
\begin{eqnarray}\nonumber
\begin{array}{lcl}
r=2:&&a_n=1,1,3,13,73,501,4051,...\\
r=3:&&a_n=1,1,4,25,211,2236,28471,...\ \ \ \ \ \ \ \ ,\ {\rm etc.}
\end{array}
\end{eqnarray}
These sequences enumerate $r$-ary forests
\cite{Sloane},\cite{Stanley},\cite{Flajolet}.}

\item[b)]{ For $q(x)=x\ln(ex)$ and $v(x)=0$ (implying $g(\lambda,x)=1$)
we have $T(\lambda,x)=e^{e^\lambda-1}x^{e^\lambda }$. This
corresponds to
\begin{eqnarray}\nonumber
{\cal N}\left[e^{\lambda a^\dag\ln(ea^\dag)a}\right] \equiv\ :
\exp\left[\left(e^{e^\lambda-1}(a^\dag)^{e^\lambda}-1\right)a\right]:\
,
\end{eqnarray}
whose coherent state matrix element with $z^{'}=1$ is equal to
\begin{eqnarray}\nonumber
\langle 1|z\rangle^{-1}\langle 1|e^{\lambda a^\dag\ln(ea^\dag)a}|z\rangle  = \exp\left[z\left(e^{e^\lambda-1}-1\right)\right]\ ,
\end{eqnarray}
which for $z=1$ generates $a_n=1,1,3,12,60,385,2471,...$
enumerating partitions of partitions \cite{Stanley},
\cite{Sloane}, \cite{Flajolet}.}
\end{itemize}
The following two examples  refer to the reverse procedure, see
Eqs.(\ref{111})-(\ref{4}). We choose first a Sheffer-type
exponential generating function and deduce $q(x)$ and $v(x)$
associated with it.

\begin{itemize}
\item[c)]{ $A(\lambda)=\frac{1}{1-\lambda}$, $B(\lambda)=\lambda$, see
Eq.(\ref{egfsheffer}). This exponential generating function for
$z=1$ counts the number of arrangements
$a_n=n!\sum_{k=0}^n\frac{1}{k!}=1,2,5,65,326,1957,...$ of the set
of $n$ elements \cite{Comtet}. The solutions of Eqs.(\ref{111}) and
(\ref{2}) are: $q(x)=1$ and $v(x)=\frac{1}{2-x}$. In terms of
bosons it corresponds to
\begin{eqnarray}\nonumber
{\cal N}\left[e^{\lambda \left(a+\frac{1}{2-a^\dag}\right)}\right]
\equiv\ :\frac{2-a^\dag}{2-a^\dag-\lambda}e^{\lambda a}:\
=\frac{2-a^\dag}{2-a^\dag-\lambda}e^{\lambda a}.
\end{eqnarray}}

\item[d)]{ For $A(\lambda)=1$ and $B(\lambda)=1-\sqrt{1-2\lambda}$ one
gets the exponential generating function of the Bessel polynomials \cite{Grosswald}. For $z=1$
they enumerate  special paths on a lattice \cite{Pittman}. The
corresponding sequence is $a_n=1,1,7,37,266,2431,...\ $. The
solutions of Eqs.(\ref{111}) and (\ref{2}) are: $q(x)=\frac{1}{2-x}$
and $v(x)=0$. It corresponds to
\begin{eqnarray}\nonumber
{\cal N}\left[e^{\lambda \frac{1}{2-a^\dag}a}\right] \equiv\
:e^{\left(1-\sqrt{(2-a^\dag)-2\lambda}\right)a}:\
\end{eqnarray}
in the boson formalism.}
\end{itemize}
These examples show that any combinatorial structure which can be
described by a Sheffer-type exponential generating function can be
cast in  boson language. This gives rise to a large number of
formulas of the above type which put them in a quantum mechanical
setting.

\section{Conclusions}
We have solved the boson normal ordering problem for the powers and exponentials of
the operators linear either in the creation or in the annihilation operator, 
{\it i.e.} $(q(a^\dag) a+v(a^\dag))^n$ and $e^{\lambda (q(a^\dag) a+v(a^\dag))}$
where $q(x)$ and $v(x)$ are arbitrary functions. This was done by
the use of umbral calculus methods \cite{Roman} in finding
representations of the monomiality principle ({\it i.e.}
representations of the Heisenberg-Weyl algebra in the space of
polynomials) and application of the coherent state methodology.
Moreover, we have established one-to-one connection between this class of
normal ordering problems and the family of Sheffer-type
polynomials and provided a wealth of combinatorial examples.

\section*{References}

\bibliographystyle{unsrt}
\bibliography{thesis}

\end{document}